\documentclass[12pt]{elsart}

\usepackage{eepic}
\usepackage{epsfig}
\def\sgn{\mbox{sgn}}
\def\cal{}
\begin{document}
\begin{frontmatter}
\title{New Stability Results for Long--Wavelength Convection Patterns}
\author[City]{Anne C.\ Skeldon} and
\author[NW]{Mary Silber}
\address[City]{Department of Mathematics, City University,
Northampton Square, London, EC1V 0HB UK, England}
\address[NW]{Department of Engineering Sciences
and Applied Mathematics, Northwestern University,
Evanston, IL 60208 USA} 

\begin{abstract}

We consider the transition from a spatially uniform state
to a steady, spatially-periodic pattern in a partial 
differential equation describing long-wavelength convection
\cite{knobloch}. 
This both extends existing work on the study of rolls,
squares and hexagons and demonstrates how 
recent generic results for the stability of
spatially-periodic patterns
may be applied in practice.  We find that squares,  
even if stable to roll perturbations, are often 
unstable when a wider class of perturbations is considered.  
We also find scenarios where transitions from hexagons 
to rectangles can occur.  In some cases we find that, near onset,
more exotic spatially-periodic planforms are preferred over the
usual rolls, squares and hexagons.

\end{abstract}
\end{frontmatter}


\section{Introduction}
\label{sec-intro}

Pattern forming instabilities arise in a wide number of physical and
chemical problems.  Model partial differential equations are used to
try to capture the essential features of the observed transitions.  In
many interesting examples such as Rayleigh-B{\'e}nard convection and
reaction-diffusion problems, the model equations are invariant under
all translations, rotations and reflections in the plane and patterns
arise at a transition from a trivial solution consisting of no
pattern.  Linear stability analysis of the trivial solution leads to a
critical curve describing how the wavenumber for instability, $k$,
depends on a parameter, $\mu$, in the problem.  For parameter
values below the critical curve the trivial solution is stable.  At a
critical parameter, $\mu_c$, instability onsets at a critical
wavenumber $k_c$.

Weakly nonlinear analysis is often used to try to predict the 
type of patterns observed once the trivial solution becomes
unstable.  Two aspects make this particularly difficult: firstly,
the rotational invariance of the problem
means that instability to a single wavenumber gives instability to
a whole circle of wavevectors.  In other words, if the trivial solution
is unstable to rolls then it is unstable to rolls with any orientation in
the plane.  Secondly, for $\mu>\mu_c$, not just a single wavenumber
but a whole band of wavenumbers is unstable.
Often this second problem is
addressed by assuming that, sufficiently close to $\mu_c$,
boundaries in any real problem will select out one particular
wavenumber and modes with neighbouring wavenumbers will be suppressed.
In the case of the first problem, 
a tacit assumption is often made that, for a given
wavenumber, only a finite number of critical
wavevectors are relevant.  
For example, four critical wavevectors oriented at $\frac{\pi}{2}$ to each
other (see figure 1(a)) are chosen  or six critical 
wavevectors oriented at $\frac{\pi}{3}$ to each other are 
chosen (see figure 1(b)).  In both cases the critical 
wavevectors generate a periodic lattice of points, 
a square lattice in the first case and a 
hexagonal lattice in the second.  Consequently,
the circle of critical wavevectors is replaced by a finite set
and a finite dimensional centre manifold exists for
the problem, of dimension four in the case of squares 
and of dimension six in the case of hexagons.  
If critical wavevectors
are used which do not generate a periodic lattice then there
is no reason {\it a priori} why a finite dimensional
centre manifold exists, since modes arbitrarily close to 
critical occur.  While non-periodic cases
have been considered
\cite{MNT}, their validity requires an additional
assumption on the suppression of these near critical modes.  

Weakly nonlinear analysis using wavevectors on a square 
lattice or a hexagonal lattice as shown in figure 1, provide
a framework for examining the relative stability of either
squares and rolls or hexagons and rolls respectively. 
In both cases generic bifurcation equations have been derived
using symmetry arguments \cite{S,BG,GKS}. 
A more complete stability analysis for rolls has been 
performed, for example by Brattkus and Davis \cite{BD}, 
who consider the relative
stability of two sets of rolls oriented at an arbitrary
angle for a problem arising in crystal growth.
Similarly, a more complete analysis can be performed for 
squares and hexagons by considering families of different
square and hexagonal lattices (for specific examples see
figure 2).  This problem has a high
degree of symmetry and using group theoretic arguments, Dionne
and Golubitsky show that, for each lattice, additional branches
other than hexagons, rolls or squares bifurcate 
as primary bifurcations \cite{Dionne,DG}. 
In spite of the fact that the Fourier transform of the new
patterns involves only one critical wavenumber, 
in physical space they appear to have more than one lengthscale.
Such ``superlattice'' patterns have recently been observed in the
Faraday crispation experiment \cite{Gollub,SP}. (For 
examples see figures 4 and 8 below.)

Using symmetry arguments, Dionne {\it et al.} derive the generic 
bifurcation equations for the families of square and 
hexagonal lattices and examine the stability of certain
primary bifurcation branches 
in terms of the coefficients of the bifurcation
equations \cite{DSS}.  This stability analysis enables two types
of statement to be made.
Firstly, since each lattice problem corresponds to a subspace of
the original unbounded problem,  and since hexagons
and squares each exist on a whole family of
lattices, the stability of these planforms 
can be considered to a countably infinite number of perturbations.
While this is 
not equivalent to completely determining the stability of 
squares and hexagons in an unbounded domain, it does 
considerably extend previous results.
Secondly, each individual lattice corresponds to
either a square or hexagonal domain
with periodic boundary conditions. For each lattice,
the relative stability of the primary branches known 
to exist from \cite{Dionne} can be calculated.  
These results, contained in \cite{DSS}, 
have not as yet been applied to any specific partial 
differential equation and it is this issue we address 
here.

In this paper,
we re-examine the relative stability of spatially-periodic
solutions to a partial differential equation
considered by Knobloch \cite{knobloch}. 
This equation, 
\begin{eqnarray}
\label{eq:longwave}
f_t & = & \alpha f-\mu\nabla^2f-\nabla^4f
              +\kappa\nabla\cdot|\nabla f|^2\nabla f \nonumber \\
&  & \qquad \qquad \qquad \qquad \qquad
   +\beta \nabla\cdot\nabla^2f\nabla f-\gamma\nabla\cdot f\nabla f
+\delta\nabla^2|\nabla f|^2,
\end{eqnarray}
describes a number of long-wavelength partial differential equations
which arise in convection problems.  For example, 
when $\kappa=1, \beta=\delta=\gamma=0$
we recover the planform equation for convection in
a layer between two poorly conducting boundaries \cite{proctor}, 
and when
$\kappa=1,\gamma=0,\beta=-\frac{\surd 7}{8}, 
\delta=-\frac{3\surd7}{8}$ equation (\ref{eq:longwave})
models long-wavelength Marangoni convection \cite{SS}.  Further 
examples are given in \cite{knobloch}.
For equation (\ref{eq:longwave}) we demonstrate, that
with relatively little additional analysis, we can derive all the
coefficients necessary to  apply the 
results from \cite{DSS}.  We thereby significantly extend 
Knobloch's stability results by inclusion of the additional
perturbations. 

In section~\ref{sec-prelim} we define the critical modes
which generate square and hexagonal lattices 
used here and the resulting generic bifurcation equations. 
The derivation of the coefficients of the bifurcation equations
is given in section~\ref{sec-analysis}.
Then in section~\ref{sec-stability}
we discuss the results for two specific cases. In Case I
we take $\gamma=0, \kappa=+1$. Knobloch called this Case B, 
the nature of our conclusions for his Case A 
are similar and we do not present them in detail. 
For Case I, provided $\beta\ne\delta$, 
the coefficient of the quadratic term
in the bifurcation equations for the hexagonal lattices is 
nonzero. This quadratic term 
renders all of the primary solution branches for the
hexagonal lattices unstable at
bifurcation \cite{ig} and thus we restrict our attention to the square
lattice bifurcation problems.  We divide our discussion
into two parts: in section~\ref{sec-stability-sq-unbound} 
we consider the stability
of squares and rolls in an unbounded domain by considering
their stability on the whole family of square lattices; in 
section~\ref{sec-stability-sq-pbc} we consider the 
particular example of long-wavelength Marangoni convection 
and show that different bifurcation scenarios can 
occur for different square lattices.
In Case II we consider ${\gamma}/{k^2} = \delta-\beta,\
\kappa=+1$. This choice of parameters yields a degenerate bifurcation
problem for the hexagonal lattice since the coefficient of the
quadratic term in the bifurcation equations is zero. 
Stable primary branches are therefore a possibility for 
all lattices and we consider both  square and hexagonal types.
We first discuss what can be deduced of
the stability of rolls, hexagons and squares
in an unbounded domain in section~\ref{sec-stability-hex-unbound};
then in section~\ref{sec-stability-hex-unfold} we discuss the 
unfolding expected if the coefficient of the quadratic
term is non-zero but 
sufficiently small. Finally, in \ref{sec-stability-hex-maran}
we discuss the specific example of Marangoni convection for
different hexagonal lattices.  Our conclusions are 
summarised in section~\ref{sec-conclude}.

\section{Preliminaries}
\label{sec-prelim}

In order to apply the analysis given in \cite{DSS}, 
we consider sets of eight or twelve critical modes whose 
wavevectors generate square or hexagonal lattices respectively,
where the length of the critical wavevector is greater than the 
distance between
neighbouring points on the lattice.  A typical example is shown in
figure 2(a) for the square lattice. This figure should be
contrasted with figure 1(a), which shows the wavevectors
which are used conventionally in pattern selection studies.  
In both figures, a circle
representing the critical wavevectors for the original unbounded
problem has been superimposed on the lattice.  In figure 1(a)
this circle only intersects the lattice at 
four points and there are consequently four critical modes,
whereas in figure 2(a) the circle intersects
the lattice at eight points and hence there are eight critical modes.  
Sufficiently close to the critical
value of the parameter, $\mu_c$, all other modes, represented by
vectors not of length $k_c$, will be
damped.  
A family of finer and finer lattices can be constructed 
each with eight points on the critical circle.  Each
lattice can be encoded by a pair of integers, $(m,n)$; for
example, the lattice shown in figure 2(a) corresponds to the case $(2,1)$
i.e. {$\bf K_{1_s}$} is 
two squares of the lattice across and one up.  The eight
wavevectors consist of two sets of four wavevectors, 
$({\bf \pm K_{1_s}, \pm K_{2_s}})$ and
$({\bf \pm K_{3_s}, \pm K_{4_s}})$, that comprise squares
and are rotated by an angle $\theta_s$ relative to each other. 
An alternative way to specify each lattice is therefore
through the lattice angle $\theta_s$, where,
\begin{equation}
\label{eq:sqtheta}
\theta_s=\cos^{-1}\left ({2mn\over m^2+n^2}\right ),
\end{equation}
and $m>n>0$ are relatively prime positive integers that are not both
odd.   Reducing the circle of critical wavevectors to 
four critical wavevectors, as shown in figure 1(a), is equivalent
to changing the original unbounded domain to a box whose side
length is 
$\frac{1}{k_c}$ and applying periodic boundary conditions.   
Using eight critical wavevectors, illustrated in
figure 2(a) for the case $(m,n)=(2,1)$, corresponds to changing
the domain to a box of 
side length $\frac{\sqrt{m^2+n^2}}{k_c}$ and again applying
periodic boundary conditions.   

In a similar way, a family of hexagonal lattices exists where the
number of critical wavevectors is twelve. 
These twelve wavevectors consist of two sets of six wavevectors,
$({\bf \pm K_{1_h},\pm K_{2_h},\pm K_{3_h}})$ and 
$({\bf \pm K_{4_h},\pm K_{5_h},\pm K_{6_h}})$, that comprise 
hexagons rotated at the angle $\theta_h$ given by
\begin{equation}
\label{eq:hextheta}
\theta_h=\cos^{-1}\left ({m^2+2mn-2n^2\over 2(m^2-mn+n^2)}\right ),
\end{equation}
where now $m>n>\frac{m}{2}$ are relatively prime positive integers,
and where $m+n$ is not a multiple of 3.  An example of the
case $(m,n)=(3,2)$ is shown in figure 2(b).
For both the square
and hexagonal lattices the requirement that $m$ and $n$ are
positive integers ensures that
the critical wavevectors generate a periodic lattice. This is 
necessary if the centre manifold theorem is to be invoked to
formally justify the use of finite-dimensional
bifurcation equations.
The remaining conditions on $m$ and $n$ ensure that 
each lattice angle corresponds to a genuinely
different case \cite{DG}. 

For the square case, letting $z_j$ be the complex amplitude of mode 
$e^{i({\bf K}_{j_s}\cdot {\bf r}) }$ where ${\bf r} = (x,y)$ then
the generic bifurcation equations take the form
\begin{eqnarray}
\label{eq:nfsq}
\dot z_1=\lambda
z_1+(a_1|z_1|^2+a_2|z_2|^2+a_3|z_3|^2+a_4|z_4|^2)z_1+{\cal O}(|{\bf z}|^5), 
 \nonumber \\
\dot z_2=\lambda
z_2+(a_2|z_1|^2+a_1|z_2|^2+a_4|z_3|^2+a_3|z_4|^2)z_1+{\cal O}(|{\bf z}|^5),
 \nonumber \\
\dot z_3=\lambda
z_3+(a_3|z_1|^2+a_4|z_2|^2+a_1|z_3|^2+a_2|z_4|^2)z_1+{\cal O}(|{\bf z}|^5),
 \\
\dot z_4=\lambda
z_4+(a_4|z_1|^2+a_3|z_2|^2+a_2|z_3|^2+a_1|z_4|^2)z_1+{\cal O}(|{\bf z}|^5).
\nonumber
\end{eqnarray}
One recovers the bifurcation equations associated with 
the wavevectors
given in figure 1 by restricting to the subspace $z_3=z_4=0$.
Equations (\ref{eq:nfsq}) have six known types of primary branch which
are listed in table 1 along with their stability assignments in
terms of the coefficients.  Note that rolls and squares are the 
same on all lattices, but the rhombs (rectangles),
super squares and anti-squares take a different form depending on
$(m,n)$. For example, changing $(m,n)$ changes the aspect ratio
of the rhombs.  An example of 
one of the super square solutions can
be seen in figure 4.  Further
examples of the different planforms are given in \cite{Dionne,DG,DSS}.

\begin{table}
\centering
\caption{Signs of eigenvalues for primary bifurcation branches on the
square lattice; $a_1,\dots,a_4$ are coefficients in the
bifurcation equation (\protect\ref{eq:nfsq}).} 
\label{square_lattice}
\begin{tabular}{|l|l|}
\hline
Planform & Signs of non-zero eigenvalues \\
\hline 
\small Rolls (R)  &  $\sgn(a_1), \quad \sgn(a_2 - a_1), 

        \quad \sgn(a_3 - a_1), \quad \sgn(a_4 - a_1)$     \\ 
${\bf z}=A_R(1,0,0,0)$  & \\
\hline
Simple Squares (S) 
       &  $\sgn(a_1 + a_2), \quad  \sgn(a_1 - a_2), \quad
          \sgn(a_3 + a_4 - a_1 - a_2)$ \\
${\bf z}=A_S(1,1,0,0)$ & \\
\hline
Rhombs (Rh$_{s1,m,n}$) 
       &  $\sgn(a_1 + a_3), \quad \sgn(a_1 - a_3), \quad
         \sgn(a_2 + a_4 - a_1 - a_3)$  \\
${\bf z}=A_{Rh}(\theta_s)(1,0,1,0)$& \\
\hline
Rhombs (Rh$_{s2,m,n}$) 
       &  $\sgn(a_1 + a_4), \quad \sgn(a_1 - a_4), \quad     
         \sgn(a_2 + a_3 - a_1 - a_4)$  \\ 
${\bf z}=A_{Rh}\bigl(\theta_s+{\pi\over 2}\bigr)(1,0,0,1)$& \\
\hline
Super Squares (SS$_{m,n}$)
       &  $ \sgn(a_1 + a_2 + a_3 + a_4), \quad
         \sgn(a_1 + a_2 - a_3 - a_4)$   \\          
${\bf z}=A_{SS}(1,1,1,1)$&  $\sgn(a_1 - a_2 + a_3 - a_4), \quad
         \sgn(a_1 - a_2 - a_3 + a_4)$   \\
 & $\sgn(\mu_0)$, where $\mu_0={\cal O}\bigl(A_{SS}^{2(m+n-1)}\bigr)$\\
\hline
Anti--Squares (AS$_{m,n}$)
       & same as super squares, except $\mu_0\to -\mu_0$\\ 
${\bf z}=A_{AS}(1,1,-1,-1)$
       & \\
\hline
\end{tabular}
\end{table}

For the hexagonal lattice problem, 
letting $z_j$ be the complex amplitude of mode 
$e^{i({\bf K}_{j_h}\cdot {\bf r}) }$ then
the generic bifurcation equations take the form
\begin{eqnarray}
\label{eq:nfhex}
\dot z_1 & = & \lambda
z_1+\epsilon \bar z_2\bar z_3 \nonumber \\
& & +(b_1|z_1|^2+b_2|z_2|^2+b_2|z_3|^2+b_4|z_4|^2
+b_5|z_5|^2+b_6|z_6|^2)z_1+{\cal O}(|{\bf z}|^4), \nonumber \\
\dot z_2 & = & \lambda
z_2+\epsilon \bar z_3\bar z_1 \nonumber \\
& & +(b_2|z_1|^2+b_1|z_2|^2+b_2|z_3|^2+b_6|z_4|^2
+b_4|z_5|^2+b_5|z_6|^2)z_2+{\cal O}(|{\bf z}|^4), \nonumber \\
\dot z_3 & = & \lambda
z_3+\epsilon \bar z_1\bar z_2 \nonumber \\
& & +(b_2|z_1|^2+b_2|z_2|^2+b_1|z_3|^2+b_5|z_4|^2
+b_6|z_5|^2+b_4|z_6|^2)z_3+{\cal O}(|{\bf z}|^4), \nonumber \\
\dot z_4 & = & \lambda
z_4+\epsilon \bar z_6\bar z_5 \\
& & +(b_4|z_1|^2+b_5|z_2|^2+b_6|z_3|^2+b_1|z_4|^2
+b_2|z_5|^2+b_2|z_6|^2)z_4+{\cal O}(|{\bf z}|^4), \nonumber \\
\dot z_5 & = & \lambda
z_5+\epsilon \bar z_4\bar z_6 \nonumber \\
& & +(b_5|z_1|^2+b_4|z_2|^2+b_6|z_3|^2+b_2|z_4|^2
+b_1|z_5|^2+b_2|z_6|^2)z_5+{\cal O}(|{\bf z}|^4), \nonumber\\
\dot z_6 & = & \lambda
z_6+\epsilon \bar z_5\bar z_4 \nonumber \\
& & +(b_6|z_1|^2+b_5|z_2|^2+b_4|z_3|^2+b_2|z_4|^2
+b_2|z_5|^2+b_1|z_6|^2)z_6+{\cal O}(|{\bf z}|^4).  \nonumber
\end{eqnarray}
The standard hexagonal bifurcation problem is recovered 
by restricting to the subspace $z_4=z_5=z_6=0$.  
Primary branches for equations (\ref{eq:nfhex}) 
are listed in table 2.  Note that
rolls and hexagons are the same on all lattices, but the rhombs and 
the super hexagons take a different form depending on $(m,n)$. 
Examples of one of the rhombs and one of the super hexagon
states are given in figure 6(b) and figure 8 respectively. As for the
square lattice, further examples of the different 
planforms may be seen in \cite{Dionne,DG,DSS}.

\begin{table}
\caption{Branching equations and signs of eigenvalues for 
primary bifurcation branches on the hexagonal lattice; $\epsilon,
b_1,\dots,b_6$ are coefficients in the bifurcation equation
(\protect\ref{eq:nfhex}).}
\label{hexagon_lattice}
\begin{tabular}{|l|l|}
\hline
Planform and branching equation	     &  Signs of non-zero eigenvalues \\ 
\hline
\small Rolls (R)  
       & $\sgn(b_1), \quad \sgn(\epsilon A_R+(b_2-b_1)A_R^2),$  \\
${\bf z}=A_R(1,0,0,0,0,0)$ 
       & $\sgn(-\epsilon A_R+(b_2-b_1)A_R^2)$, \\
$0=\lambda A_R +b_1 A_R^3+{\cal O}(A_R^5)$
       & $\sgn(b_4 - b_1), \quad \sgn(b_5 - b_1), \quad \sgn(b_6 - b_1).$\\
\hline
Simple Hexagons (H$^\pm$)
       & $ \sgn(\epsilon A_H+2(b_1+2b_2)A_H^2),$ \\
${\bf z}=A_H(1,1,1,0,0,0)$ 
       &  $ \sgn(-\epsilon A_H+(b_1-b_2)A_H^2), $     \\ 
$0=\lambda A_H +\epsilon A_H^2 \qquad \qquad $
       & $ \sgn(-\epsilon A_H+(b_4+b_5+b_6-b_1-2b_2)A_H^2)$,  \\ 
$\qquad \qquad + (b_1 + 2b_2) A_H^3 + {\cal O}(A_H^4)$  
       & $ \sgn(-\epsilon A_H+{\cal O}(A_H^3)).$          \\ 
\hline
Rhombs (Rh$_{h1,m,n}$) 
       & $\sgn(b_1 + b_4), \quad \sgn(b_1 - b_4)$                    \\ 
${\bf z}=A_{Rh}(\theta_h)(1,0,0,1,0,0)$ 
       & $\sgn(\mu_1),\quad \sgn(\mu_2),\quad$  where,                          \\
$0=\lambda A_{Rh} + (b_1+b_4) A_{Rh}^3+{\cal O}(A_{Rh}^5)$
       & $\mu_1+\mu_2=(-2b_1-2b_4+2b_2+b_5+b_6)A_{Rh}^2$,             \\ 
       & $\mu_1\mu_2=
-\epsilon^2A_{Rh}^2+(b_1+b_4-b_2-b_5)(b_1+b_4-b_2-b_6)A_{Rh}^4.$      \\ 
\hline
Rhombs (Rh$_{h2,m,n}$)     
       & $\sgn(b_1 + b_5), \quad \sgn(b_1 - b_5)$                   \\
${\bf z}=A_{Rh}\bigl(\theta_h+{2\pi\over 3}\bigr)(1,0,0,0,1,0)$
       & $\sgn(\mu_1),\quad \sgn(\mu_2), \quad$ where,                           \\
$0=\lambda A_{Rh} + (b_1+b_5) A_{Rh}^3+{\cal O}(A_{Rh}^5)$
       & $\mu_1+\mu_2=(-2b_1-2b_5+2b_2+b_4+b_6)A_{Rh}^2,$            \\  
       & $\mu_1\mu_2=
-\epsilon^2A_{Rh}^2+(b_1+b_5-b_2-b_4)(b_1+b_5-b_2-b_6)A_{Rh}^4.$     \\ 
\hline
Rhombs (Rh$_{h3,m,n}$)     
       & $\sgn(b_1 + b_6), \quad \sgn(b_1 - b_6)$                   \\ 
${\bf z}=A_{Rh}\bigl(\theta_h-{2\pi\over 3}\bigr)(1,0,0,0,0,1)$
       & $\sgn(\mu_1),\quad \sgn(\mu_2), \quad $ where,                           \\
$0=\lambda A_{Rh} + (b_1+b_6) A_{Rh}^3+{\cal O}(A_{Rh}^5)$
       & $\mu_1+\mu_2=(-2b_1-2b_6+2b_2+b_4+b_5)A_{Rh}^2,$            \\ 
       & $\mu_1\mu_2=-\epsilon^2A_{Rh}^2
+(b_1+b_6-b_2-b_4)(b_1+b_6-b_2-b_5)A_{Rh}^4$,                         \\ 
\hline
       & $\sgn(\epsilon A_{SH}+2(b_1+2b_2+b_4+b_5+b_6)A_{SH}^2)$ \\
Super Hexagons (SH$^\pm_{m,n})^{**}$ 
       & $\sgn(\epsilon A_{SH}+2(b_1+2b_2-b_4-b_5-b_6)A_{SH}^2)$ \\ 
${\bf z}=A_{SH}(1,1,1,1,1,1)$ 
       & $\sgn(-\epsilon A_{SH}+{\cal
O}(A_{SH}^3))$, $\sgn(-\epsilon A_{SH} +{\cal O}( A_{SH}^3))^*$\\
$0=\lambda A_{SH}+\epsilon A_{SH}^2+(b_1+2b_2)A_{SH}^3$ 
       & $\sgn(\mu_1),\quad \sgn(\mu_2),\quad $ where,       \\
$\quad+(b_4+b_5+b_6)A_{SH}^3+{\cal O}(A_{SH}^4)$ 
       & $\mu_1+\mu_2=-4\epsilon A_{SH} +4(b_1-b_2)A_{SH}^2,$ \\ 
       & $\mu_1\mu_2=4(\epsilon A_{SH}-(b_1-b_2)A_{SH}^2)^2$ \\
       &
 $\quad \qquad\qquad-2((b_4-b_5)^2+(b_4-b_6)^2+(b_5-b_6)^2))A_{SH}^4$\\
       & $\sgn(\mu_0)$, where, $\mu_0={\cal O}(A_{SH}^{2(m-1)}).$\\ 
\hline
\multicolumn{2}{l}{$^*$ These two eigenvalues differ at ${\cal O}(A_{SH}^3)$.}\\ 
\multicolumn{2}{l}{$^{**}$ Super triangles \protect\cite{SP} have the same 
eigenvalues except sgn$(\mu_0) \to$ -sgn$(\mu_0)$.} \\
\end{tabular}
\end{table}

\section{Calculation of the Coefficients}
\label{sec-analysis}

The form of the bifurcation equations (\ref{eq:nfsq}) 
and (\ref{eq:nfhex}) is determined by the symmetry of the
problem.  However,  the coefficients, and therefore the
stability of the different planforms, depend upon the
particular application.  One important result in \cite{DSS} is
that, although high order terms in the bifurcation
equations are required to find the relative stability of
some planforms, much is fixed by the cubic order truncation.
Therefore, in this section we determine the 
coefficients of the quadratic and cubic terms in the bifurcation
equations (\ref{eq:nfsq}) and (\ref{eq:nfhex}) from the
long-wavelength equation (\ref{eq:longwave}).   
Our approach is to
infer these quantities from the branching equations for rolls, simple
squares, rhombs, and simple hexagons, which are computed using
perturbation theory as follows.
We let
\begin{eqnarray*}
f & = & \epsilon f_0 + \epsilon^2 f_1 + \epsilon^3 f_2 + \ldots \nonumber \\
\mu & = & \mu_c + \epsilon \mu_1 + \epsilon \mu_2 + \ldots,
\end{eqnarray*}
where $\mu$ is the bifurcation parameter and instability onsets
at the critical value $\mu_c=2(-\alpha)^{1/2}$, at the critical 
wavenumber $k_c=(-\alpha)^{1/2}$. 
In turn, we take 
\begin{eqnarray}
\label{eq:ampsdef}
{\rm Rolls}:\quad f_0&=&A_Re^{ikx}+\cdots+c.c.\nonumber\\
{\rm Squares}:\quad f_0&=&A_S\bigl(e^{ikx}+e^{iky}\bigr)+\cdots+c.c.\\
{\rm Hexagons}:\quad f_0&=&A_H\bigl(e^{ikx}+e^{ik(-x+\sqrt{3}\ y)/2}+
e^{ik(-x-\sqrt{3}\ y)/2}\bigr)+\cdots+c.c.\nonumber\\
{\rm Rhombs}:\quad f_0&=&A_{Rh}\bigl(e^{ikx}+e^{ik(cx+sy)}\bigr)
+\cdots+c.c.,\ {\rm where} 
\ c\equiv\cos(\theta),\ s\equiv\sin(\theta)\ .\nonumber
\end{eqnarray}
At $O(\epsilon^3)$ the solvability condition gives
an equation for the amplitudes $A_R,A_S,A_H,A_{Rh}$
respectively. That is, 
\begin{eqnarray}
\label{eq:bifs}
\dot A_R&=&\lambda A_R  + C_R A_R^3 +{\cal O}(A_R^5),\nonumber\\
\dot A_S&=&\lambda A_S  + C_S A_S^3 +{\cal O}(A_S^5),\nonumber\\
\dot A_H&=&\lambda A_H  + \epsilon A_H^2 + C_{H} A_H^3 +{\cal O}(A_H^4),\\
\dot A_{Rh}&=&\lambda A_{Rh} + C_{Rh} A_{Rh}^3 +{\cal O}(A_{Rh}^5),\nonumber
\end{eqnarray} 
where $\lambda=k^2(\mu-\mu_c)$ and 
\begin{eqnarray}
\label{eq:coeffs}
\epsilon&=&k^4\left(\beta+{\gamma\over k^2}-\delta\right)\nonumber\\
C_R&=& -{k^4}\left(3\kappa+2\left(\beta+\delta-{\epsilon\over
3k^4}\right )\left(\delta+{\epsilon\over 3k^4}\right)\right)\nonumber\\
C_S&=&C_R-2k^4\left(\kappa+2\left(\delta^2-{\epsilon^2\over k^8}\right)
\right )\\
C_H&=&C_R-{3k^4\over 2}\left(4\kappa+\left(\beta+2\delta-{\epsilon\over
k^4}\right) \left(2\delta+{\epsilon\over k^4}\right)\right)\nonumber\\
C_{Rh}(\theta)&=&C_S-4k^4\cos^2\theta
\left (\kappa+\beta\delta-{\beta\epsilon\over k^4
(1-4\cos^2\theta)}-{8\epsilon^2(1-2\cos^2\theta)\over
k^8(1-4\cos^2\theta)^2}\right ) \ .\nonumber
\end{eqnarray}

Note that, as expected, $C_{Rh}({\pi\over 2})=C_S$. Also note that
$C_{Rh}$ diverges as $\theta\to\frac{\pi}{3}$ since   
when $\theta=\frac{\pi}{3}$ 
there is resonance between $e^{ikx}$ and $e^{ik(cx+sy)}$. 

The cubic coefficients in the equivariant bifurcation equations
(\ref{eq:nfsq}) and (\ref{eq:nfhex}) are readily expressed in terms of
the branching coefficients $C_R,C_S,C_H$ and $C_{Rh}$.  For instance,
if (\ref{eq:nfsq}) is restricted to the simple squares subspace ${\bf
z}=(A_S,A_S,0,0)$, we find 
\begin{equation}
\label{eq:sqsubsp}
\dot A_S=\lambda A_S+(a_1+a_2)A_S^3. 
\end{equation}
Comparing equation (\ref{eq:sqsubsp}) with the appropriate
branching equation from equations 
(\ref{eq:bifs}) gives $a_1+a_2=C_S$. 
Similarly, by restricting to subspaces for rolls and rhombs
for the square lattice problems (\ref{eq:nfsq}), we find
\begin{eqnarray}
\label{eq:sqcoeffs}
a_1 & = & C_R, \quad a_2=C_S-C_R, \nonumber \\
a_3 &=& C_{Rh}(\theta_s)-C_R,
             \quad a_4=C_{Rh}\Bigl(\theta_s+{\pi\over 2}\Bigr)-C_R,
\end{eqnarray}
where $\theta_s\in(0,{\pi\over 2})$ takes on one of the discrete set of
values (\ref{eq:sqtheta}).  Similarly, restriction to subspaces
for rolls, hexagons and rhombs for 
the hexagonal
lattice bifurcation problems (\ref{eq:nfhex}) leads to 
\begin{eqnarray}
\label{eq:hexcoeffs}
b_1&=&C_R,\qquad  b_2={1\over 2}(C_H-C_R),
\qquad b_4=C_{Rh}(\theta_h)-C_R,\nonumber\\ 
b_5&=&C_{Rh}\Bigl(\theta_h+{2\pi\over 3}\Bigr)-C_R,\qquad 
b_6=C_{Rh}\Bigl(\theta_h-{2\pi\over 3}\Bigr)-C_R,
\end{eqnarray}
where $\theta_h\in(0,{\pi\over 3})$ takes on one of the values in the
discrete set (\ref{eq:hextheta}). 

Note that the expressions for $C_R$, $C_S$,
$\epsilon$, and $C_H$ are given in \cite{knobloch} and, in that paper,
are used to calculate the coefficients $a_1,a_2,b_1$ and $b_2$. The
remaining expression for $C_{Rh}$ is the only additional calculation
required to enable all the remaining coefficients in the 
bifurcation equations (\ref{eq:nfsq}) and (\ref{eq:nfhex}) to be 
found.

\section{Stability Results}
\label{sec-stability}

In this section we use the bifurcation equations (\ref{eq:nfsq}) and
(\ref{eq:nfhex}) to determine the relative stability of the
steady planforms which are given in section 2.
The results depend on the parameters $\kappa,\ \beta,\ \delta,$ and
$\gamma$ in the long-wave equation (\ref{eq:longwave}). They also
depend on the size of the periodic domain through the lattice parameters
$(m,n)$.
We restrict our discussion to two cases and, where possible,
compare and contrast our results with those given in \cite{knobloch}.

The task of establishing the stability of each planform is 
considerably eased by the fact that the
eigenvalues determining the relative stability of each primary
branch are derived in terms of the coefficients of the Taylor
expansion of the bifurcation equations in \cite{DSS}. The
signs of the eigenvalues are
summarized in tables~\ref{square_lattice} and \ref{hexagon_lattice}
for the square and hexagonal lattices, respectively.
The sign of the first quantity listed for
each planform gives the branching
direction; if this eigenvalue is negative (positive), then the branch
is supercritical (subcritical).  If $\epsilon\neq0$
then simple hexagons and super hexagons
bifurcate transcritically; all other patterns arise through pitchfork
bifurcations. We distinguish between the two branches of hexagons,
denoted H$^+$ and H$^-$, which satisfy $A_H>0$ and $A_H<0$,
respectively.  Similarly, there are two distinct branches of super
hexagons, denoted by SH$^\pm_{m,n}$. 
Omitted from the tables are the zero
eigenvalues associated with translations of the patterns, and also
information about the multiplicities of each eigenvalue; this
information can be found in \cite{DSS}.  
Note that certain eigenvalues in tables~\ref{square_lattice} and
\ref{hexagon_lattice} are not determined at cubic order in the
bifurcation equations. For instance, the relative stability of super
squares and anti-squares depends on a resonant term of ${\cal
O}(|{\bf z}|^{2m+2n-1})$. However, in this case, if super squares and
anti-squares are neutrally stable at cubic order, then, generically,
exactly one of the two states is stable.  There is an analogous
stability result for super hexagons and triangles \cite{SP}.

In the case of the square lattice, the eigenvalues that
depend only on $a_1$ and $a_2$ can be determined by considering the
restricted bifurcation problem $z_3=z_4=0$ in equation (\ref{eq:nfsq}).
Similarly, those results for the hexagonal
lattice that depend only on $\epsilon$, $b_1$ and $b_2$ can be
obtained by considering the simpler hexagonal bifurcation problem.
In general, the signs of all the remaining eigenvalues are dependent
on the choice of lattice.  
However, in the special case when $\epsilon=0$, 
we find that those eigenvalues which are unchanged on
permutation of $a_3$ and $a_4$ are independent of $\theta_s$, and
those which are unchanged on permutation of $b_4,b_5$ and $b_6$ are
independent of $\theta_h$.
This is due to the particularly simple $\theta$-dependence 
of $C_{Rh}(\theta)$ in equation (\ref{eq:coeffs}) when
$\epsilon=0$.  Also, in the case of the square lattices,
results for angles close to $\frac{\pi}{6}$ and $\frac{\pi}{3}$ must be 
interpreted with care because of the singularity in $C_{Rh}(\theta)$
at $\theta=\frac{\pi}{3}$ that occurs due to resonant interactions.

In each of the cases we discuss below, we evaluate 
the signs of the eigenvalues for each planfrom and
determine if and where they change sign.  We present the results
in the form of bifurcation sets separating different
regions of stability
and instability for the relevant patterns. Note that along the
stability boundaries themselves, the bifurcation problem is 
degenerate and the bifurcation equations (\ref{eq:nfsq}) and
(\ref{eq:nfhex}) are insufficient to locally determine the
bifurcation structure.  These degenerate points have been
analysed by Knobloch \cite{knobloch}.  

\subsection{Case I: $\gamma=0,\ \kappa=+1$}
\label{sec-stability-sq}

In case I, solutions on the hexagonal lattices are unstable
at onset unless $\beta=\delta$ i.e. $\epsilon=0$ in 
equations (\ref{eq:nfhex}).  Thus here we focus on the
square lattice problems and defer discussion of the
hexagonal cases to section \ref{sec-stability-hex}.   

\subsubsection{The stability of squares and rolls}
\label{sec-stability-sq-unbound}
If we can show that squares
or rolls are unstable on any one lattice then
they must be unstable in the original unbounded problem 
sufficiently close to onset.  From the hexagonal lattice
result, we can immediately infer that rolls are unstable
to hexagonal perturbations for $\beta\neq\delta$. Here
we show that squares are also 
unstable unless $\beta$ is sufficiently close to $\delta$.
The bifurcation sets for rolls
and for squares in the $(\beta,\delta)$-plane are presented in 
figures 3(a) and (b) respectively. 

The sign of the branching eigenvalue for rolls, $\sgn
(a_1)$, is always negative, so rolls always bifurcate
supercritically.  The sign of
$a_2-a_1$ determines the relative stability of squares and rolls
and the corresponding bifurcation line, $a_1=a_2$ in figure 3(a), 
is identical to the
line $q(0)=0$ given in figure 2(b) of \cite{knobloch}. The
relative stability of rolls and the two rhombic patterns is determined
by the signs of $a_3-a_1$ and $a_4-a_1$.  
Since $a_3$ and $a_4$ are
dependent on $\theta_s$, the precise position of the
corresponding bifurcation curves $a_1=a_3$ and $a_1=a_4$ 
depends on the choice of lattice: those shown in figure 3(a)
are for the case $(m,n)=(2,1)$. Qualitatively the
picture is the same in all cases corresponding to 
$0<\theta_s<\frac{\pi}{4}$.
For $\theta_s > \frac{\pi}{4} $, the picture is 
similar except $a_3-a_1$ and $a_4-a_1$ switch roles.
The region of stability is indicated by 
the shaded wedges in the $(\beta,\gamma)$-plane
between $a_1=a_2$ and $a_1=a_4$ ($a_1=a_2$ and $a_1=a_3$ if 
$\theta_s >\frac{\pi}{4}$). As the lattice angle, $\theta_s$, 
approaches $\frac{\pi}{6}$ or $\frac{\pi}{3}$ 
the region of stability of the
rolls is reduced to a narrower and narrower region occurring
only for large $|\beta|$ and $|\delta|$. At precisely
$\theta=\frac{\pi}{3}$ (or the complementary $\frac{\pi}{6}$)
the hexagonal lattice must be considered.

In figure 3(b) we show the analogous bifurcation set for 
squares showing where each of the three expressions given in 
table 1 for the 
eigenvalues for squares change sign.  The lines $a_1=a_2$
and $a_1=-a_2$ correspond to the lines $q(0)=0$ and $p_N(0)=0$ given
by Knobloch.  In his study he found 
the squares were preferred to rolls for the region between these two curves.
However, this region is significantly reduced when instability
to super square (or anti-square) states is included through the
eigenvalue $a_3+a_4-a_1-a_2$.  The position of the 
corresponding bifurcation line 
given by $a_3+a_4=a_1+a_2$ is again dependent on the value of
the lattice angle.  As the lattice angle approaches $\frac{\pi}{6}$ or
$\frac{\pi}{3}$ the region of stability of the squares is 
reduced to a narrower
and narrower region. Interestingly this narrow region always
includes the line $\beta=\delta$, for which the hexagonal problem 
is degenerate.  Recall that, for hexagonal lattices,
it is only in this
degenerate case that stable planforms can exist at onset.
Thus, in summary, we find that there are only stable squares or rolls
when $\beta\approx\delta$, that is $\epsilon\approx0$.

Knobloch also considered the case where $\beta=\delta, \gamma \neq 0$
which he refers to as case A.
In this case we find that the region of stability 
for the rolls and for squares is 
diminished to a narrow region about $\gamma=0$
as the lattice angle approaches 
$\frac{\pi}{6}$ or $\frac{\pi}{3}$.  Since 
$\gamma=0$ corresponds to the degenerate hexagonal
problem, again 
we find that squares and rolls are unstable unless 
$\epsilon\approx 0$.

\subsubsection{Stability of other planforms: 
                 the example of Marangoni convection}
\label{sec-stability-sq-pbc}
As discussed in section~\ref{sec-prelim}, for each lattice, 
there are in fact six primary branches known to exist, 
any one of which could in principle 
be stable.  Since many of the eigenvalues
are dependent on the lattice angle $\theta_s$,
the precise region of stability for each
state is dependent on the choice of lattice, 
i.e. on the size of the domain for a box
with periodic boundary conditions.  Consequently, for a given
physical problem, which
planforms are stable at onset, can be dependent on the
size of the box.  We illustrate this with the example
of Marangoni convection which corresponds to 
$\beta=-\frac{\surd 7}{8}, \delta=-\frac{3\surd7}{4}$ 
and $\gamma=0$. This lies within 
the region where squares are preferred to rolls in Knobloch's analysis.
In contrast, we find that one of the following scenarios
occurs:
\begin{tabbing}
$15.79$  \= $< \theta_s < $ \= $18.34^o:\quad$ \= $Rh_{s2,m,n}$ stable.  \kill
$0$      \> $< \theta_s < $ \> $15.79^o:$      \>  Bistability 
of $Rh_{s2,m,n}$ and squares. e.g. (m,n)=(6,5). \\
$15.79$  \> $< \theta_s < $\> $18.34^o:$      \> $Rh_{s2,m,n}$ stable.  
e.g. (m,n)=(11,8). \\
$18.34$  \> $< \theta_s < $\> $39.26^o:$      \> Everything unstable, 
e.g. (m,n)=(2,1). \\
$39.26$  \> $< \theta_s < $\> $43.71^o:$      \> $Rh_{s2,m,n}$ stable.
e.g. (m,n)=(9,4). \\
$43.71$  \> $< \theta_s < $\> $44.67^o:$      \> Super squares or anti-squares
are stable.
e.g. (m,n)=(19,8). \\
$44.67$  \> $< \theta_s < $\> $45.0^o:$      \> Squares stable.
e.g. (m,n)=(29,12). \\
%
%
%
%
%
%
\end{tabbing}
The results for $45^0<\theta_s<90^0$ are essentially the same with
$Rh_{s1}$ and $Rh_{s2}$ interchanged.  Recall that the size of
the periodic box for which these results apply is given by
$\frac{\sqrt{m^2+n^2}}{k_c}$ and that the aspect ratio of the 
rhombs (rectangles) 
is given by $\frac{m-n}{m+n}$ for Rh$_{s1,m,n}$ and by 
$\frac{n}{m}$ for Rh$_{s2,m,n}$.
In figure 4 we show an example of 
the bifurcation diagram close to onset for the case $(m,n)=(19,8)$. 
We know that either super squares or anti-squares
are stable and have drawn the super square case.  We have
not calculated which of these two  planforms
is preferred at onset 
since this is determined at an ${\cal O} (2(m+n)-1) = {\cal O} (53)$ 
truncation of the bifurcation equations!   

\subsection{Case II: 
The degenerate case $\frac{\gamma}{k_c^2} = \delta-\beta,\ \kappa=+1$}
\label{sec-stability-hex}

In the degenerate case $\frac{\gamma}{k_c^2} = \delta-\beta$ the
quadratic coefficient, $\epsilon$, is zero in equation (\ref{eq:nfhex})
and both square and hexagonal
lattices can give locally stable planforms.  We have therefore
evaluated the signs of the 
eigenvalues listed in both table~\ref{square_lattice}
and table~\ref{hexagon_lattice} which are determined at
a cubic truncation of the bifurcation equations.  

\subsubsection{The stability of rolls, hexagons and squares}
\label{sec-stability-hex-unbound}
Our results are presented in the form 
of a bifurcation set in the $(\beta,\delta)$-plane shown
in figure 5.

We first recap the relative stability
results given in \cite{knobloch} for the subspaces
$z_4=z_5=z_6=0$ of equation (\ref{eq:nfhex}) and
$z_3=z_4=0$ of equation (\ref{eq:nfsq})
before discussing the results
of our extended analysis.  First, in the  subspace of
the hexagonal bifurcation problem it is found:

\begin{itemize}
\item Hexagons are stable in region 1.
\item Rolls are stable in regions 2,3 and 4.
\end{itemize}

In contrast, for the subspace of the square lattice
bifurcation problem:

\begin{itemize}
\item Squares are stable in region 1, 2 and 3.
\item Rolls are stable only in region 4.
\end{itemize}

Together the results for the two lattices suggest that in regions 2
and 3, for an unbounded domain, that hexagons are unstable to rolls
but that rolls are unstable to squares. However, there is
no formal way of directly studying if hexagons are unstable to
squares.

In our analysis of the finer lattices the
results given above are modified.  For a given hexagonal
lattice we find
hexagons and rhombs (Rh$_{h3,m,n}$) are stable in region 1. 
Rhombs (Rh$_{h3,m,n}$) are stable in region 2 and 
rolls are stable only in regions 3 and 4.
All other planforms are unstable.
The position of the line dividing regions 2 and 3  is
dependent on the choice of lattice.  
As the lattice angle approaches $\frac{\pi}{6}$, this line 
approaches the line $a_1=a_2$. The rhombs (rectangles) have aspect
ratio depending on the lattice angle but lying
between $\frac{1}{\surd 3}$ and 1.

For a given finer square lattice we find
the stability results are unchanged for squares and rolls, 
however rhombs can be stable in regions 1 and 2.
In particular, 
for the rhombs, we find that rhombs, $Rh_{s2,m,n}$ are stable if
$0<\theta_s<\frac{\pi}{6}$ and the rhombs, Rh$_{s1,m,n}$
are stable if $\frac{\pi}{3}<\theta_s<\frac{\pi}{2}$. These have
aspect ratio dependent on $\theta_s$, but again lying 
between $\frac{1}{\surd 3}$ and 1.  
In addition, there are regions of stable rhombs within region 1
which exist for $\frac{\pi}{6}<\theta<\sin^{-1}\frac{1}{\surd 3}\approx35.3^0$
and for $\frac{\pi}{3}>\theta>\cos^{-1}\frac{1}{\surd 3}\approx54.7^0$
(aspect ratios between
$\frac{1}{\surd 3}$ and $\frac{\surd 3 -1}{\surd 2}$).

In summary, although the regions of stability of
the hexagons and squares are unchanged by the extended bifurcation
analysis,  the inclusion of the rhombic states allows for
bistability of hexagons and rhombs of aspect ratio close
to one.  

\subsubsection{Unfolding the degenerate problem}
\label{sec-stability-hex-unfold}

When $\epsilon \neq 0$ the bifurcation equations
for the hexagonal lattices contain a quadratic term and 
all planforms are necessarily locally
unstable.  However, if the quadratic term is sufficiently
small compared with the cubic term, stable states may result
through secondary bifurcations.  
For example, in the conventional analysis of hexagons using
six critical wavevectors (figure 1(b)), for the degenerate bifurcation 
problem in regions 2,3 and 4 of figure 5, rolls and hexagons
both bifurcate as pitchforks and rolls are stable.  If
a small quadratic term is added, then local to 
the trivial solution at $\mu_c$ there are no stable
bifurcating branches and the hexagons 
bifurcate transcritically creating two distinct branches,
H$^+$ and H$^-$.  An example bifurcation diagram of 
such a scenario is shown in figure
6(a). In this figure it can be seen that
as $\mu$ is increased, there is
a hysteretic transition to hexagons from the trivial 
solution.  On further increase in $\mu$ there is
a second hysteretic transition, this time between 
hexagons and rolls.

In our extended bifurcation analysis, again all planforms
arise at pitchfork bifurcations in the degenerate case. 
When $\epsilon \neq 0$ both super hexagons and hexagons 
now bifurcate as transcritical bifurcations and all
states are locally unstable.  Secondary bifurcations
can, however, again stabilise some of the branches.  If
we consider $\beta$ and $\delta$ with values corresponding
to region 2 of figure 5 and if
$\frac{\epsilon}{(1-\cos^2 \theta_h)}$ is 
sufficiently small,  then the bifurcation 
diagram of figure 6(a) is replaced by that 
shown in figure 6(b).  The hexagons now undergo a hysteretic
transition to rhombs rather than to rolls.  Remember that
the aspect ratio of the rhombs is dependent on the lattice:
those shown
in figure 6(b) are for the lattice $(m,n)=(3,2)$ which
results in rhombs of aspect ratio $\frac{\surd 3}{2}
\approx 0.87$.    

The criterion that
 $\frac{\epsilon}{(1-\cos^2 \theta_h)}$ is sufficiently small comes
from requiring that the $\epsilon$ corrections to the 
cubic coefficients are small enough to be neglected.   If
this is not the case, which is inevitable for fixed
$|\epsilon| \ll 1$ if the full
range in $\theta_h$ is considered, then a sequence of
more complicated transitions can occur.  We illustrate
this by once again considering the case of Marangoni 
convection.  

\subsubsection{Unfolding the degenerate problem: the example of Marangoni
convection}
\label{sec-stability-hex-maran}
Although the 
Marangoni problem is nondegenerate, previous studies
\cite{SS} have assumed that the quadratic term 
is sufficiently
small and that it can legitimately be compared with the cubic
terms.  Specifically, the Marangoni problem has
$\beta=-\surd 7 /8, \delta=-3 \surd7/4, \epsilon=\frac{5\surd7}{8}$ 
and although $\epsilon$ does not appear to be small,
the cubic coefficients themselves are relatively large. For example,
$a_1=-\frac{1615}{144}$ and $a_2=-\frac{1707}{128}$
resulting in a saddle-node bifurcation on the 
hexagonal branch occurring at $\mu=-0.018$.
We have calculated all the eigenvalues that are determined
at cubic order for the Marangoni case for the family
of hexagonal lattices 
and found the corresponding bifurcation lines.  
The results are summarised in figure
7.   The horizontal axis gives the lattice angle, 
$\theta_h$. For the eigenvalues shown, this diagram is
reflection symmetric about the line $\theta_h=\frac{\pi}{6}$ and
we have therefore only shown $0<\theta_h<\frac{\pi}{6}$.
For simplicity, we have only shown 
bifurcation lines which separate
stable from unstable regions of the different 
planforms. For $\theta_h$ approximately
between $15^o$ and $45^o$ a transition from hexagons to
rhombs occurs in a similar manner to that shown in
figure 6(b).  However, for $\theta_h$ 
outside this range the $\epsilon$ corrections to the
cubic coefficients result in many further secondary
transitions of significance. 
Five main shaded regions are shown indicating different
combinations of stable planforms, including   
regions of stable super hexagons with rhombs and/or
hexagons for a cubic truncation of the bifurcation
equations (note that the relative stability of
super hexagons and triangles
is only determined at ${\cal O} (2m-1)$ \cite{SP}). 
There is, in addition, a very narrow region
to the left of the black region where super hexagons
are the only stable planform.  This region is too small
to be readily discernible from figure 7, but is apparent 
in the bifurcation diagram shown in figure 8. 
This bifurcation diagram was computed from the 
bifurcation equation (\ref{eq:nfhex}) for the
lattice $(m,n)=(12,7)$, i.e. $\theta_h=10.99^o$.  
Only branches which are stable for $\mu$ close to critical
are drawn.  In this particular case there is a region
of bistability between hexagons and super hexagons.  
A hysteretic transition between hexagons and super hexagons
can occur.  An illustration of the super hexagon
state for the case $(m,n)=(12,7)$ is also shown.  Although
such complex patterns have not yet been seen in convection
problems, recent experimental results of the Faraday 
crispation experiments do show such ``superlattice 
patterns'' \cite{Gollub}.

Note that, if all lattices (all $\theta_h$) are considered, 
then hexagons
are necessarily unstable (see $0<\theta_h<8^o$ in figure 7). 
However, in this case where we have fixed $\epsilon\neq0$,
it is unclear whether this local problem is still valid
for the amplitudes where secondary bifurcations arise.
Nevertheless, it demonstrates some of the intriguing 
possibilities associated with this bifurcation problem.

\section{Conclusions}
\label{sec-conclude}

Standard low-dimensional bifurcation
analyses of squares and hexagons give only a 
restricted stability analysis of these planforms.  
We have shown that, with one additional perturbation
calculation, i.e. the calculation of $C_{Rh}$,  all
the additional coefficients required to apply the extended
stability analysis of \cite{DSS} are determined.  We have 
performed this calculation for the case of the 
long-wavelength convection equation (\ref{eq:longwave}) and
analysed the results in two main cases.  In Case I
we found that  extending the stability results 
significantly increased the known region of instability
for squares in an unbounded domain.  In particular, 
we find that squares are only stable at onset if 
$\epsilon\approx 0$.  This is interesting given that it is
already known that all solutions which are periodic on
a hexagonal lattice are unstable at onset unless
$\epsilon=0$. 
For a given box with periodic boundary conditions we found
that the predicted planform at onset was strongly dependent
on the size of the box: some of the more exotic planforms
such as super squares and anti-squares could be stable.
In Case II we showed that regions of bistability of 
rhombs and hexagons exist.  This gave a formal setting
for studying the transition between hexagons and rhombs
of aspect ratio close to 1 (although not squares).  

\ack
The research of MS was supported by NSF grant DMS-9404266 and by an
NSF CAREER award DMS-9502266.  ACS thanks the Nuffield Foundation 
and the Royal Society for their support.
The authors also appreciate the
hospitality of the Hat Creek Radio Observatory, where much 
of the work for this paper was done.

\newpage

\begin{figure}
\setlength{\unitlength}{0.3pt}
\begin{picture}(1007,900)(-150,0)
\thicklines
\put(530,456){\ellipse{438}{438}}
\multiput(96,19)(0,218.5){5}{\blacken\ellipse{15}{15}}
\multiput(313,19)(0,218.5){5}{\blacken\ellipse{15}{15}}
\multiput(530,19)(0,218.5){5}{\blacken\ellipse{15}{15}}
\multiput(746,19)(0,218.5){5}{\blacken\ellipse{15}{15}}
\multiput(963,19)(0,218.5){5}{\blacken\ellipse{15}{15}}
\put(530,456){\vector(1,0){215}}
\put(530,456){\vector(0,1){215}}
\put(530,456){\vector(-1,0){215}}
\put(530,456){\vector(0,-1){215}}
\put(630,410){${\bf k_1}$}
\put(545,550){${\bf k_2}$}
\put(-100,960){\bf \Large (a)} 
\end{picture}

\vspace*{70pt}

\setlength{\unitlength}{0.3pt}
\begin{picture}(1007,900)(-150,0)
\thicklines
\put(530,457){\ellipse{438}{438}}
\multiput(154,19)(0,219){5}{\blacken\ellipse{15}{15}}
\multiput(342,128)(0,219){4}{\blacken\ellipse{15}{15}}
\multiput(530,19)(0,219){5}{\blacken\ellipse{15}{15}}
\multiput(717,128)(0,219){4}{\blacken\ellipse{15}{15}}
\multiput(905,19)(0,219){5}{\blacken\ellipse{15}{15}}
\path(530,456)(530,665)
\path(523,640)(530,665)(537,640)(523,640)
\path(530,456)(708,352)
\path(690,370)(708,352)(683,358)(690,370)
\path(530,456)(346,352)
\path(371,358)(346,352)(364,370)(371,358)
\path(530,456)(530,247)
\path(523,272)(530,247)(537,272)(523,272)
\path(530,456)(708,560)
\path(690,542)(708,560)(683,554)(690,542)
\path(530,456)(346,560)
\path(371,554)(346,560)(364,542)(371,554)
\put(545,550){${\bf k_1}$}
\put(630,410){${\bf k_2}$}
\put(435,360){${\bf k_3}$}
\put(-100,960){\bf \Large (b)}
\end{picture}
\caption{(a) Square lattice generated by four wavevectors on
the critical circle oriented at $\frac{\pi}{2}$ to each other.
(b) Hexagonal lattice generated by six wavevectors on
the critical circle oriented at $\frac{\pi}{3}$ to each other.}
\end{figure}

\newpage

\begin{figure}
\setlength{\unitlength}{0.30pt}
\begin{picture}(1007,900)(-150,0)
\thicklines
\put(530,456){\ellipse{438}{438}}
\multiput(142,65)(0,97.75){9}{\blacken\ellipse{15}{15}}
\multiput(239,65)(0,97.75){9}{\blacken\ellipse{15}{15}}
\multiput(336,65)(0,97.75){9}{\blacken\ellipse{15}{15}}
\multiput(433,65)(0,97.75){9}{\blacken\ellipse{15}{15}}
\multiput(530,65)(0,97.75){9}{\blacken\ellipse{15}{15}}
\multiput(626,65)(0,97.75){9}{\blacken\ellipse{15}{15}}
\multiput(723,65)(0,97.75){9}{\blacken\ellipse{15}{15}}
\multiput(820,65)(0,97.75){9}{\blacken\ellipse{15}{15}}
\multiput(917,65)(0,97.75){9}{\blacken\ellipse{15}{15}}
\put(530,456){\vector(2,1){192}}
\put(530,456){\vector(1,2){96}}
\put(530,456){\vector(-1,2){96}}
\put(530,456){\vector(-2,1){192}}
\put(530,456){\vector(2,-1){192}}
\put(530,456){\vector(1,-2){96}}
\put(530,456){\vector(-1,-2){96}}
\put(530,456){\vector(-2,-1){192}}
\put(733,564){${\bf K_{1_s}}$}
\put(636,661){${\bf K_{3_s}}$}
\put(363,671){${\bf K_{2_s}}$}
\put(266,574){${\bf K_{4_s}}$}
\put(-100,910){\bf \Large (a)}
\put(570,500){\bf$\theta_s$}
\path(620,501)(612.4,513.7)(607,520.7)(601.1,527.1)(594.7,533.1)(587.7,538.4)(575,546)
\end{picture}

\vspace*{36pt}

\setlength{\unitlength}{0.3000pt}
\begin{picture}(1007,900)(-150,0)
\thicklines
\put(530,456){\ellipse{438}{438}}
\multiput(175,167)(0,82.5){8}{\blacken\ellipse{15}{15}}
\multiput(246,126)(0,82.5){9}{\blacken\ellipse{15}{15}}
\multiput(317,167)(0,82.5){8}{\blacken\ellipse{15}{15}}
\multiput(388,126)(0,82.5){9}{\blacken\ellipse{15}{15}}
\multiput(459,167)(0,82.5){8}{\blacken\ellipse{15}{15}}
\multiput(530,126)(0,82.5){9}{\blacken\ellipse{15}{15}}
\multiput(600,167)(0,82.5){8}{\blacken\ellipse{15}{15}}
\multiput(671,126)(0,82.5){9}{\blacken\ellipse{15}{15}}
\multiput(742,167)(0,82.5){8}{\blacken\ellipse{15}{15}}
\multiput(813,126)(0,82.5){9}{\blacken\ellipse{15}{15}}
\multiput(884,167)(0,82.5){8}{\blacken\ellipse{15}{15}}
\path(530,456)(671,621)
\path(642,599)(664,613)(653,590)(642,599)
\path(530,456)(327,495)
\path(353,497)(327,495)(350,483)(353,497)
\path(530,456)(597,259)
\path(582,280)(597,259)(595,285)(582,280)
\path(530,456)(597,652)
\path(582,631)(597,652)(595,626)(582,631)
\path(530,456)(327,417)
\path(353,415)(327,417)(350,429)(353,415)
\path(530,456)(666,297)
\path(643,314)(666,297)(653,323)(643,314)
\path(530,456)(389,621)
\path(418,599)(396,613)(407,590)(418,599)
\path(530,456)(733,495)
\path(707,497)(733,495)(710,483)(707,497)
\path(530,456)(463,259)
\path(478,280)(463,259)(465,285)(478,280)
\path(530,456)(463,652)
\path(478,631)(463,652)(465,626)(478,631)
\path(530,456)(733,417)
\path(707,415)(733,417)(710,429)(707,415)
\path(530,456)(394,297)
\path(417,314)(394,297)(407,323)(417,314)
\put(680,615){${\bf K_{1_h}}$}
\put(255,520){${\bf K_{2_h}}$}
\put(580,200){${\bf K_{3_h}}$}
\put(250,390){${\bf K_{5_h}}$}
\put(590,675){${\bf K_{4_h}}$}
\put(690,275){${\bf K_{6_h}}$}
\put(-100,880){\bf \Large (b)}
\put(565,540){\bf$\theta_h$}
\path(611.5,550.7)(601.7,558.4)(592.5,564.3)(582.8,569.3)(570.7,574.2)
\end{picture}
\caption{(a) Square lattice generated by eight wavevectors on the 
critical circle.  This case shows $(m,n)=(2,1)$.
(b) Hexagonal lattice generated by twelve wavevectors on the critical
circle.  This case shows $(m,n)=(3,2)$.}
\end{figure}

\newpage
 
\begin{figure}
\setlength{\unitlength}{0.9pt}
\begin{picture}(100,425)(-50,-300)
\epsfig{file=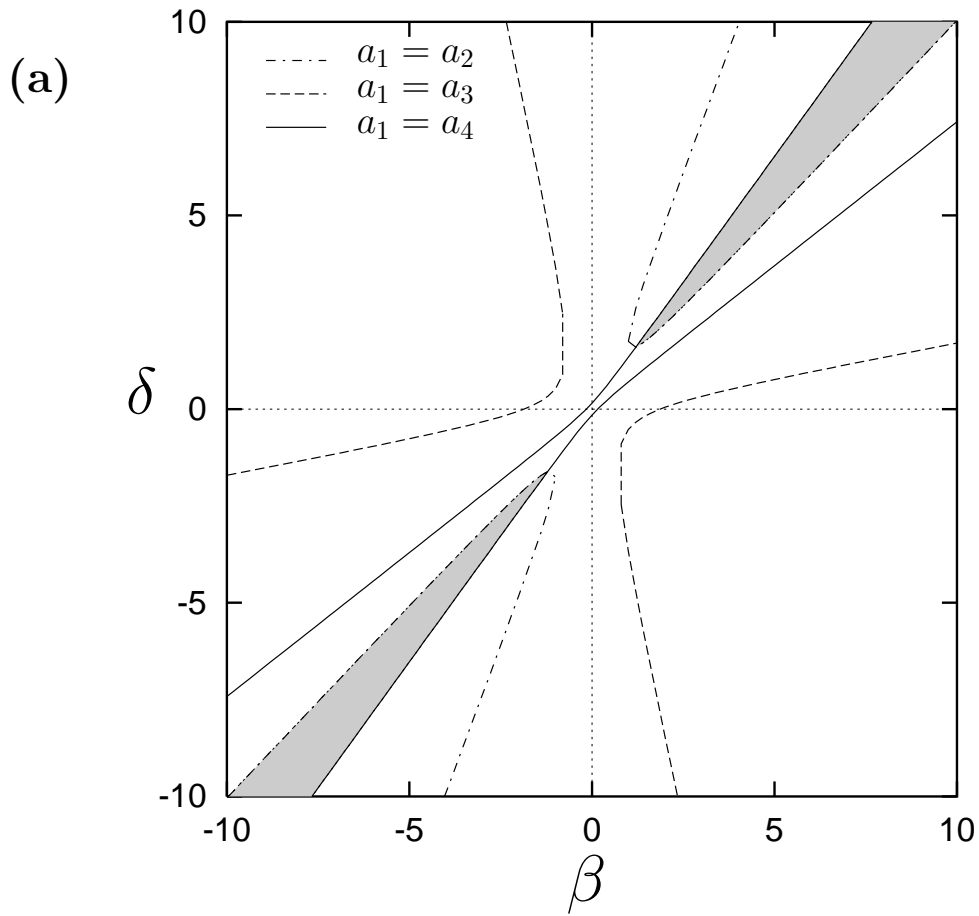}
\put(-265,-190){\huge $\beta$}
\put(-450,15){\huge $\delta$}
\put(-353,163){\large $a_1=a_2$}
\put(-353,148){\large $a_1=a_3$}
\put(-353,133){\large $a_1=a_4$}
\put(-500,150){\bf \Large (a)}
\end{picture}

\vspace*{150pt}

\begin{picture}(100,100)(-50,-180)
\epsfig{file=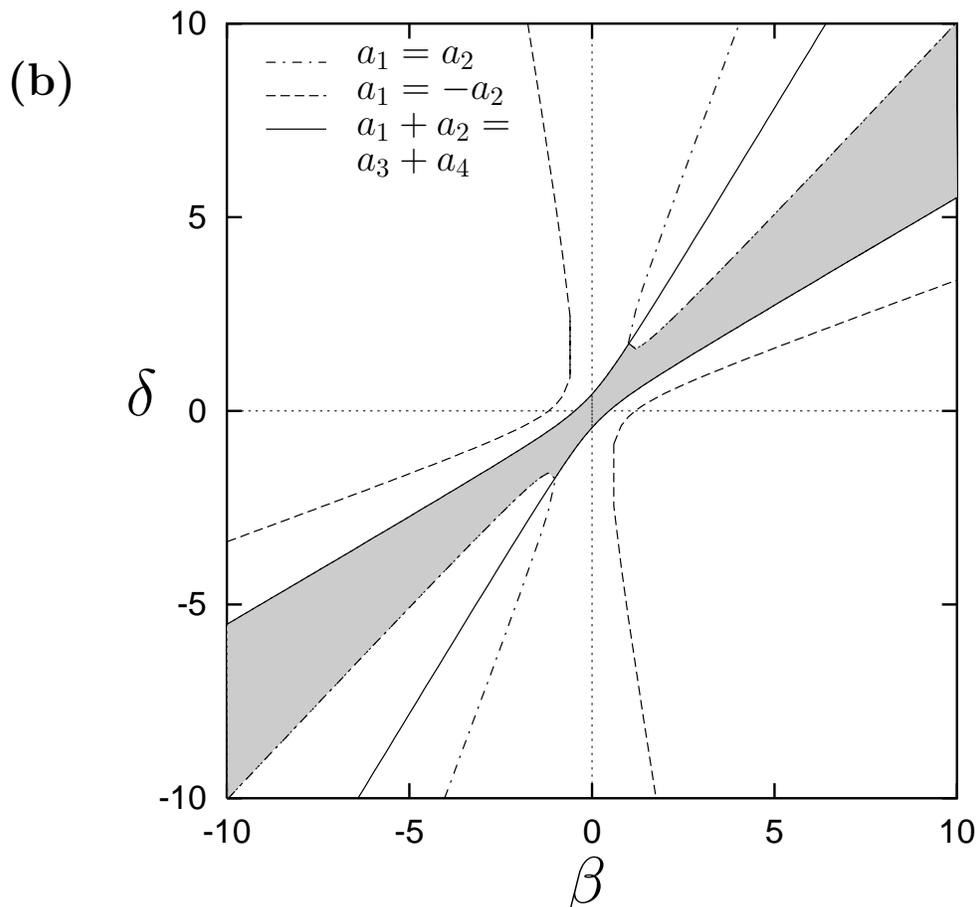}
\put(-265,-190){\huge $\beta$}
\put(-450,15){\huge $\delta$}
\put(-353,163){\large $a_1=a_2$}
\put(-353,148){\large $a_1=-a_2$}
\put(-353,133){\large $a_1+a_2=$}
\put(-353,118){\large$a_3+a_4$}
\put(-500,150){\bf \Large (b)}
\end{picture}
\caption{Bifurcation sets in case I for $(m,n)=(2,1)$.  The lines
indicate transitions in the number of negative eigenvalues.
(a) Bifurcation set for rolls.  The shaded region shows the 
region where rolls are preferred.
(b) Bifurcation set for squares.  The shaded region shows the 
region where squares are preferred.}
\end{figure}

\clearpage
\newpage

\begin{figure} 

\setlength{\unitlength}{1pt}
\begin{picture}(200,500)(-40,0)
\epsfig{file=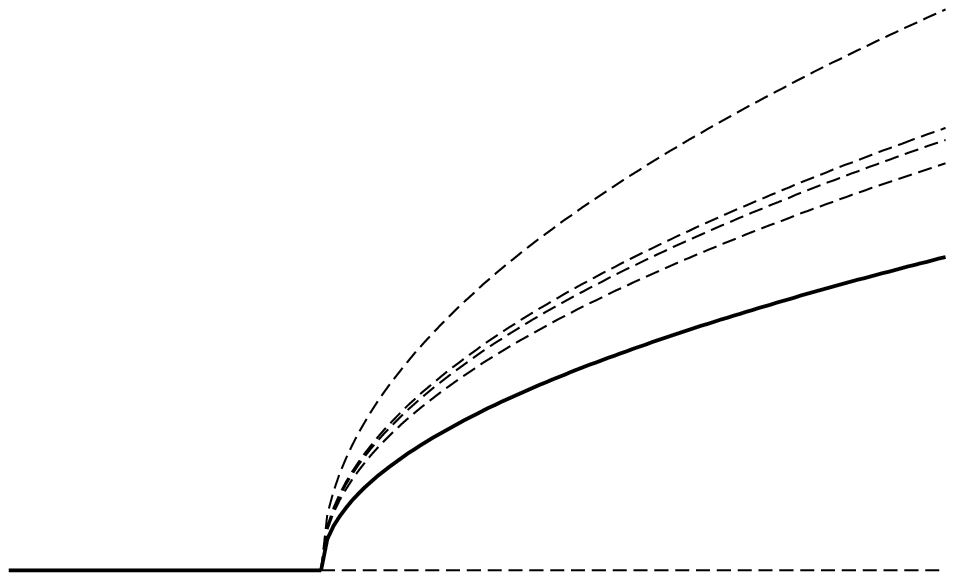}
\put(-500,185){\epsfig{file=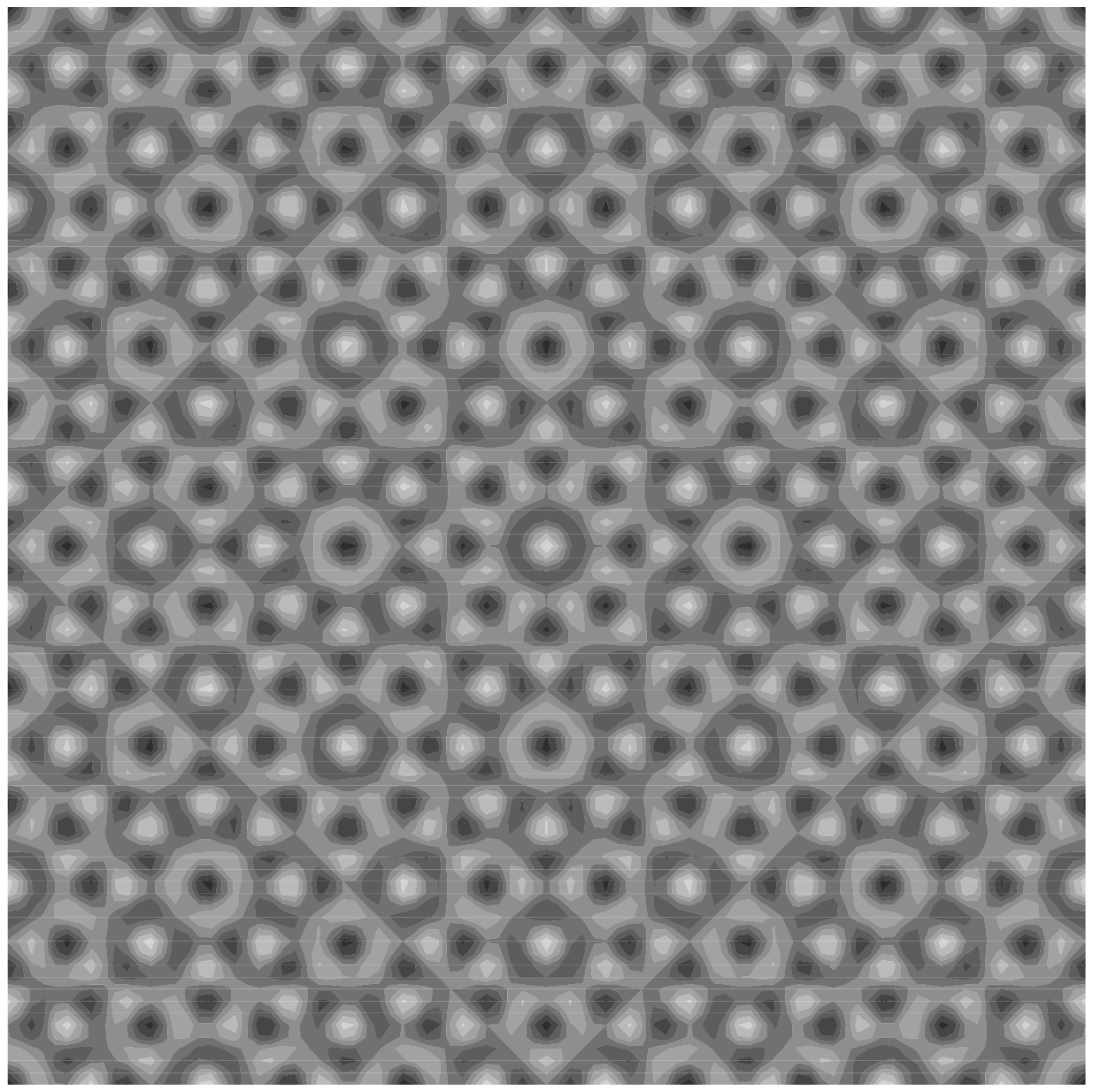}}
\put(-200,210){\vector(1,-1){72}}
\put(-290,35){\Large$\mu$}
\put(-20,160){\large SS$_{19,8}$ and AS$_{19,8}$}
\put(-20,182){\large Rh$_{s1,19,8}$}
\put(-20,195){\large S}
\put(-20,208){\large Rh$_{s2,19,8}$}
\put(-20,235){\large R}
\put(-320,60){\vector(1,0){30}}
\put(-320,60){\vector(0,1){30}}
\put(-100,380){\Large (or anti-squares)}
\put(-207,72){$\bullet$}
\put(-207,62){\large$\mu_c$}
\end{picture}

\caption{Bifurcation diagram for the square lattice $(m,n)=(19,8)$ 
for the
Marangoni problem, $\beta=-\frac{\surd 7}{8}, \delta=-\frac{3\surd 7}{4}$.
The stable branch is marked with a solid line and
the unstable branches with dashed lines.  An example of
the super square planform is shown: this is spatially periodic, i.e.
not a quasipattern. Only one period of the pattern is
shown.}
\end{figure}

\clearpage
\newpage
 
\begin{figure}

\setlength{\unitlength}{1pt}
\begin{picture}(100,340)(-30,-210)
\epsfig{file=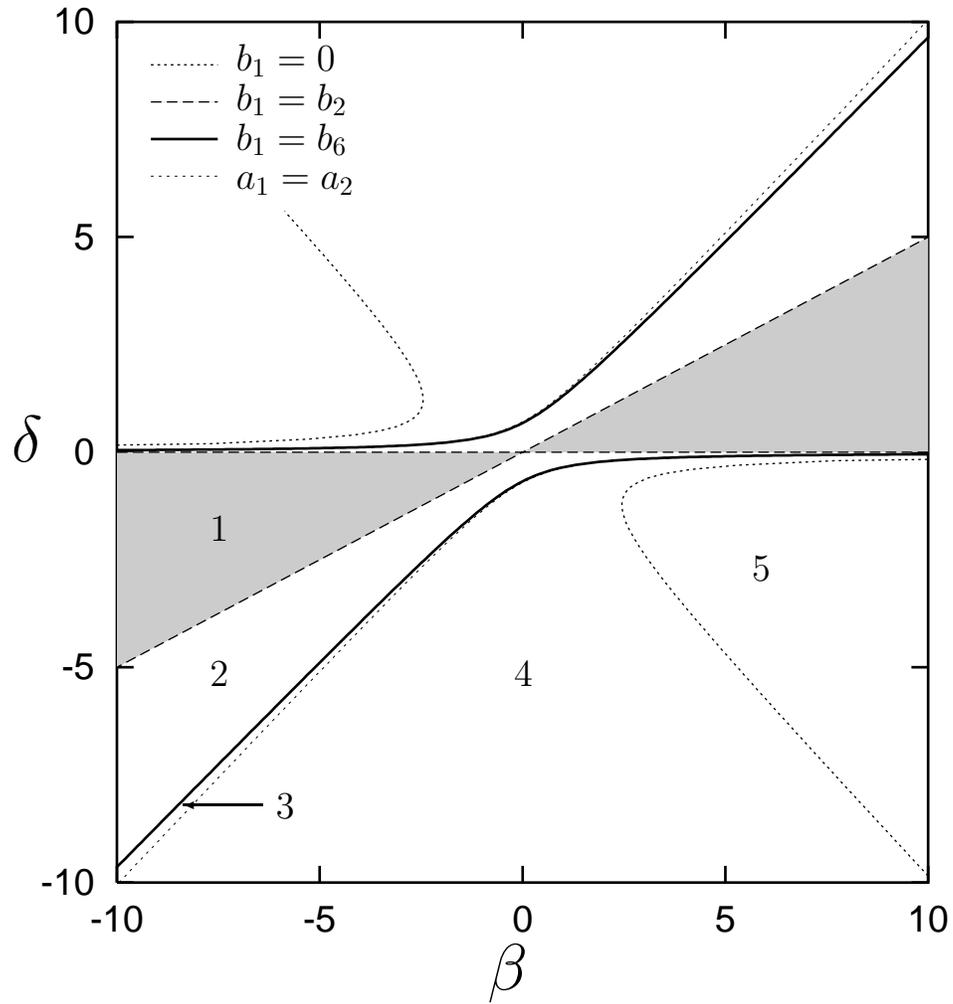}
\put(-220,-170){\huge $\beta$}
\put(-400,30){\huge $\delta$}
\put(-315,178){\large $b_1=0$}
\put(-315,163){\large $b_1=b_2$}
\put(-315,148){\large $b_1=b_6$}
\put(-315,133){\large $a_1=a_2$}
\put(-325,0){\large 1}
\put(-325,-55){\large 2}
\put(-300,-105){\large 3}
\put(-210,-55){\large 4}
\put(-120,-15){\large 5}
\put(-305,-100){\vector(-1,0){30}}
\end{picture}

\caption{Bifurcation set in the case $\epsilon=0$.  Only
critical eigenvalues which divide stable from unstable planforms
are shown. The different regions are discussed in the text.}
\end{figure}

\clearpage
\newpage
 
\begin{figure}

\begin{picture}(200,405)(-50,-110)
\epsfig{file=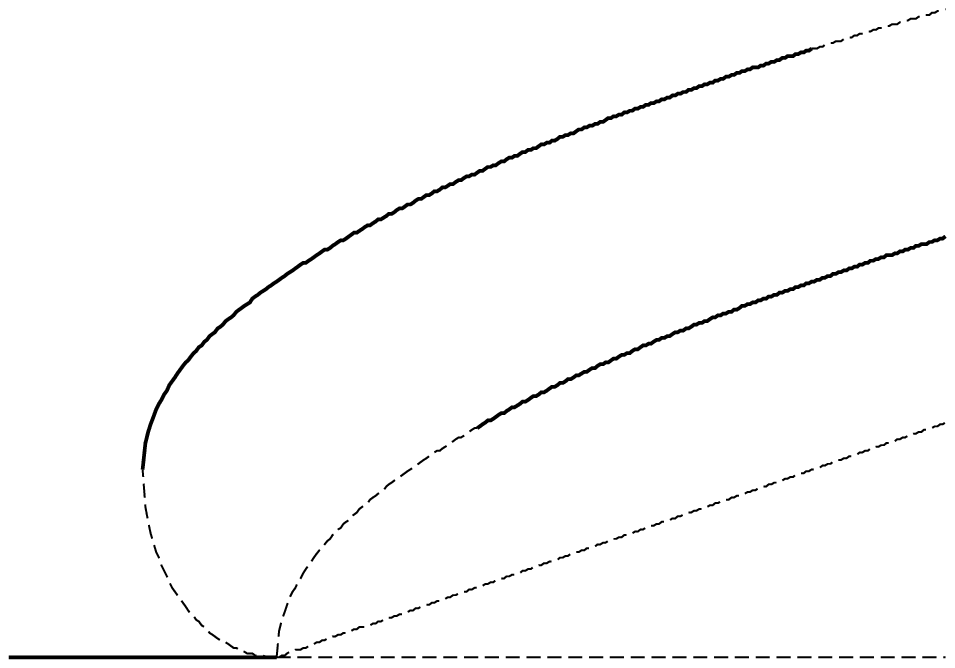}
\put(-350,-295){\epsfig{file=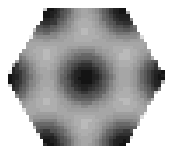}}
\put(-600,-180){\epsfig{file=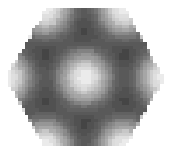}}
\put(-415,-100){\epsfig{file=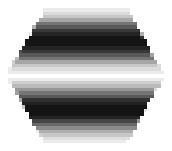}}
\put(-126,270){\vector(0,-1){105}}
\put(-280,210){\vector(2,-1){50}}
\put(-70,112){\vector(-2,1){15}}
\put(-290,35){\Large$\mu$}
\put(-20,140){\large H$_-$}
\put(-20,195){\large R}
\put(-20,260){\large H$+$}
\put(-320,60){\vector(1,0){30}}
\put(-320,60){\vector(0,1){30}}
\put(-220,72.5){$\bullet$}
\put(-258,124.5){$\bullet$}
\put(-163,138){$\bullet$}
\put(-62,249){$\bullet$}
\put(-220,62.5){$\mu_c$}
\put(-420,310){\bf \Large (a)}
\end{picture}

\begin{picture}(200,200)(-50,40)
\epsfig{file=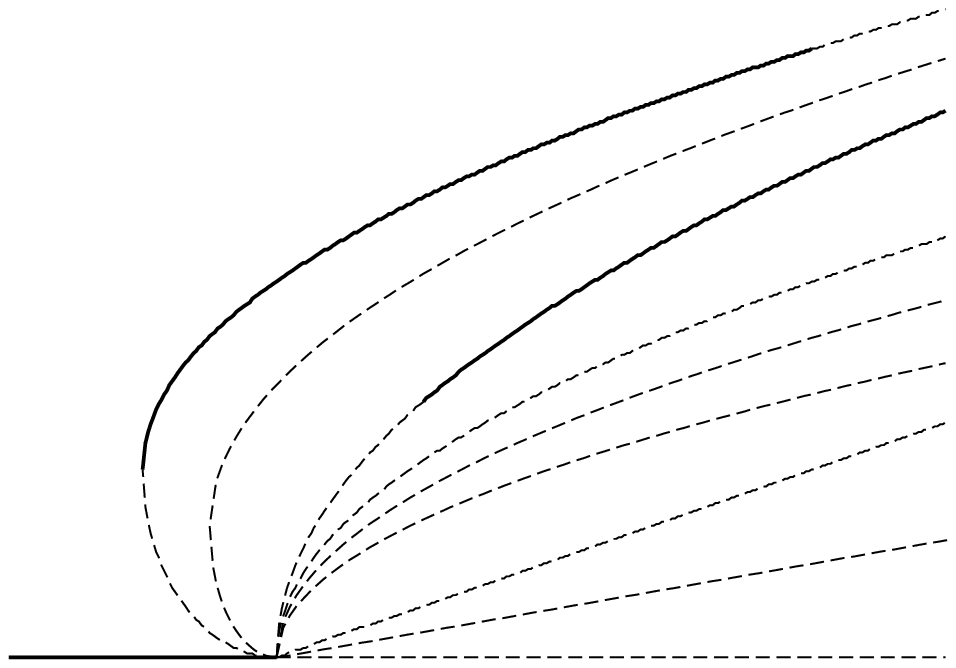}
\put(-610,-155){\epsfig{file=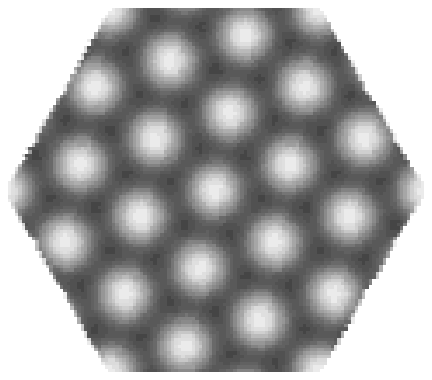}}
\put(-398,-60){\epsfig{file=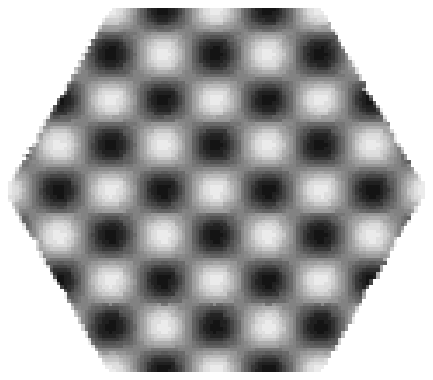}}
\put(-90,275){\vector(0,-1){68}}
\put(-265,205){\vector(2,-1){40}}
\put(-290,35){\Large$\mu$}
\put(-20,140){\large H$^-$}
\put(-20,195){\large R}
\put(-20,260){\large H$^+$}
\put(-20,106){\large SH$^-_{3,2}$}
\put(-20,245){\large SH$^+_{3,2}$}
\put(-20,156){\large Rh$_{h1,3,2}$}
\put(-20,176){\large Rh$_{h2,3,2}$}
\put(-20,230){\large Rh$_{h3,3,2}$}
\put(-320,60){\vector(1,0){30}}
\put(-320,60){\vector(0,1){30}}
\put(-220,72.5){$\bullet$}
\put(-258,124.5){$\bullet$}
\put(-163,138){$\bullet$}
\put(-62,249){$\bullet$}
\put(-180,144){$\bullet$}
\put(-220,62.5){$\mu_c$}
\put(-420,360){\bf \Large (b)}
\end{picture}
\caption{Unfoldings of the bifurcation diagrams relevant to
region 2 in figure 5 for $\epsilon$ sufficiently small. 
(a) The bifurcation diagram obtained by considering the
subspace $z_4=z_5=z_6=0$ of equation (\ref{eq:nfhex}).
(b) The new bifurcation diagram when the additional branches
in our extended analysis is considered.  Note that the
rhombic (rectangular) pattern has an aspect ratio close
to 1.} 
\end{figure}

\clearpage
\newpage
 
\begin{figure}
\begin{picture}(200,450)(-20,-50)
\put(-50,-100){\epsfig{file=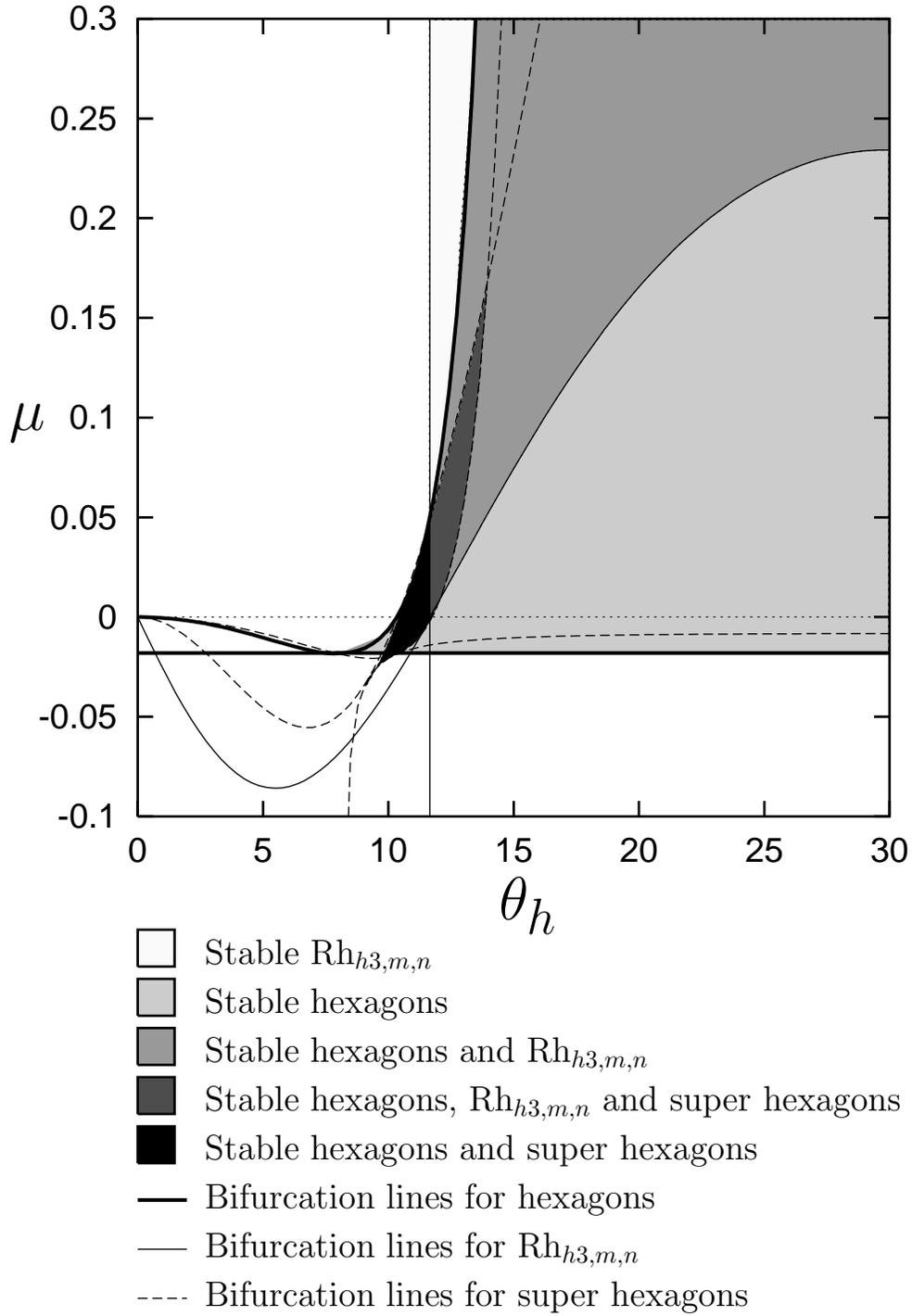}}
\put(195,130){\Huge$\theta_h$}
\put(-5,330){\Huge$\mu$}
\put(75,112){\large Stable Rh$_{h3,m,n}$}
\put(75,92){\large Stable hexagons}
\put(75,72){\large Stable hexagons and Rh$_{h3,m,n}$}
\put(75,52){\large Stable hexagons, Rh$_{h3,m,n}$ and super hexagons}
\put(75,32){\large Stable hexagons and super hexagons}
\put(75,12){\large Bifurcation lines for hexagons}
\put(75,-8){\large Bifurcation lines for Rh$_{h3,m,n}$}
\put(75,-28){\large Bifurcation lines for super hexagons}
\end{picture}

\caption{Bifurcation set for the
Marangoni problem, $\beta=-\frac{\surd 7}{8}, 
\delta=-\frac{3\surd 7}{4}, \epsilon=\frac{5\surd 7}{8}$ 
as a function  of the lattice angle $\theta_h$.}
\end{figure}

\clearpage
\newpage
 
\begin{figure}

\begin{picture}(200,635)(-440,-35)
\put(-540,220){\epsfig{file=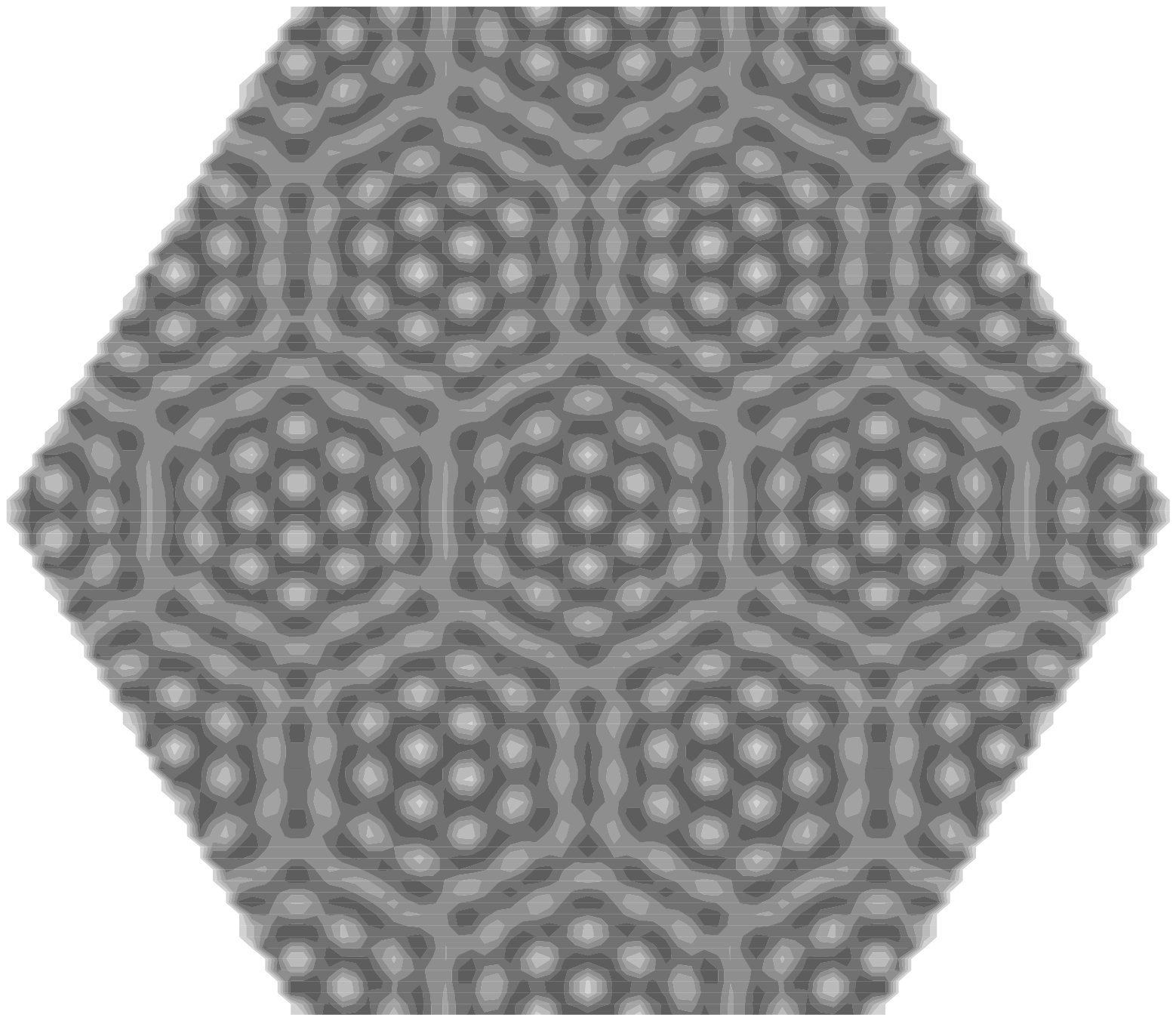}}
\put(-397,0){\epsfig{file=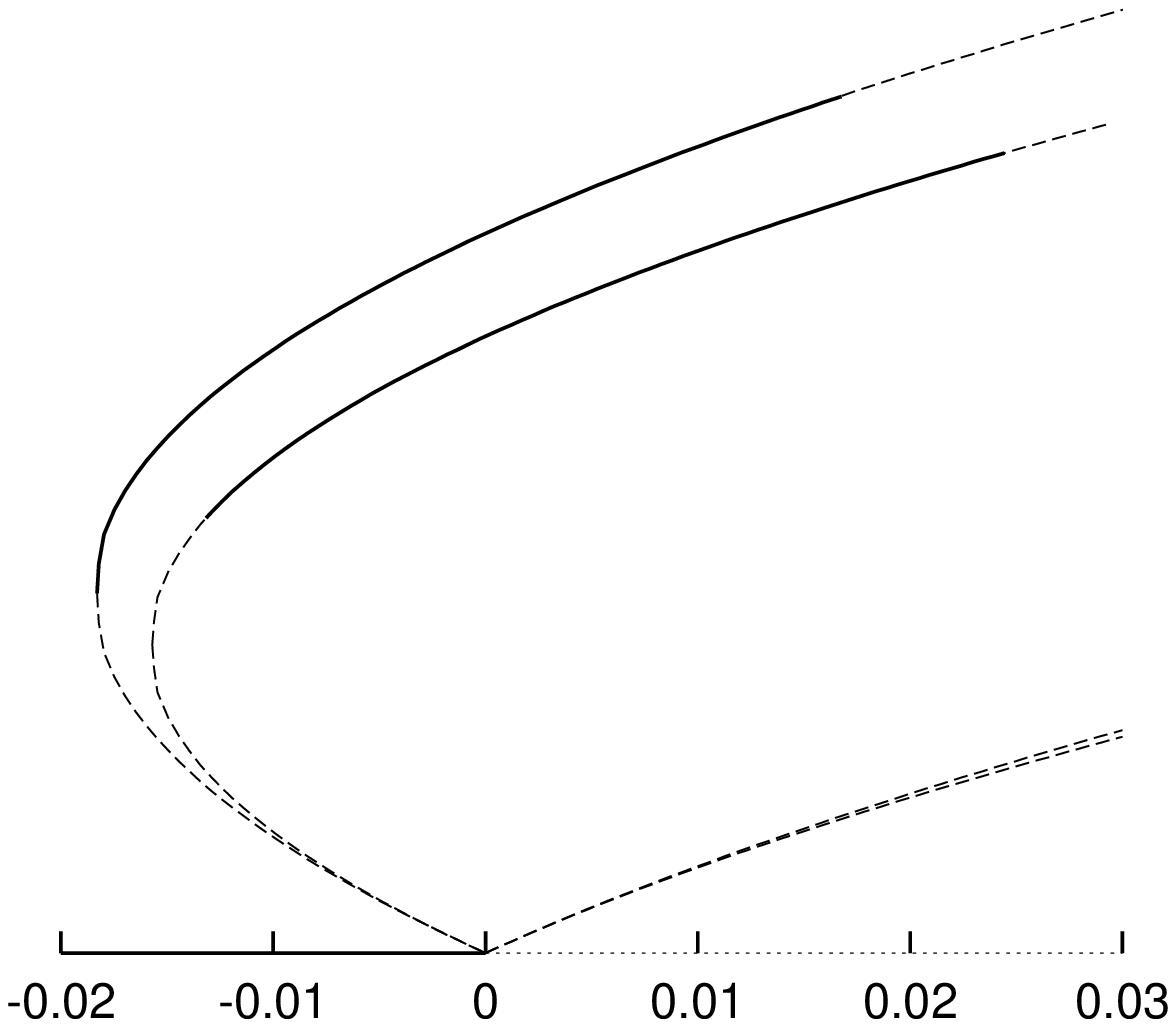}}
\put(-268,253){\vector(1,-1){53}}
\put(-210,-25){\Large$\mu$}
\put(-20,290){\large H$^+$}
\put(-20,90){\large H$^-$}
\put(-20,75){\large SuH$^-_{12,7}$}
\put(-20,255){\large SuH$^+_{12,7}$}
\put(-210,18){$\bullet$}
\put(-322,123){$\bullet$}
\put(-306,109){$\bullet$}
\put(-293,142){$\bullet$}
\put(-107,266){$\bullet$}
\put(-65,248){$\bullet$}
\end{picture}
\caption{Example bifurcation diagram for the Marangoni 
problem in a box with periodic boundary conditions. This
case is for $(m,n)=(12,7)$.  Only branches which can
be stable are shown.}  
\end{figure}


\newpage
\begin{thebibliography}{99}
\bibitem{knobloch} E.\ Knobloch. 
Pattern selection in long-wavelength convection. 
{\it Physica D} {\bf 41} (1990) 450--479.

\bibitem{MNT} B.A.\ Malomed, A.A.\ Nepomnyashchii and
M.I.\ Tribelskii.
Two-dimensional quasiperiodic structures in nonequilibrium
systems.
{\it Sov. Phys. JETP} {\bf 69} (1989) 388--396.

\bibitem{S} J.W.\ Swift.
Bifurcation and symmetry in convection
{\it PHD Thesis} (1984) University of California, Berkeley.

\bibitem{BG} E.\ Buzano and M.\ Golubitsky.
Bifurcation on the hexagonal lattice and the planar B{\'e}nard
problem.
{\it Philos. Trans. R. Soc. A} {\bf 308} (1983) 617--667.

\bibitem{GKS} M.\ Golubitsky, E.\ Knobloch and J.W.\ Swift.
Symmetries and pattern selection in Rayleigh-B{\'e}nard convection.
{\it Physica D} {\bf 10} (1984) 249--276.


\bibitem{BD}K.\ Brattkus and S.H.\ Davis.
Cellular growth near absolute stability
{\it Phys.\ Rev.\ B} {\bf 38} (1988) 11452--60.

\bibitem{Gollub}  A.\ Kudrolli, B.\ Pier and J.P.\ Gollub.
Superlattice patterns in surface waves.
{\it preprint} (1997).

\bibitem{SP} M.\ Silber and M.R.E.\ Proctor.
Nonlinear competition between small and large hexagonal patterns.
{\it preprint} (1997).

\bibitem{Dionne} B.\ Dionne. Spatially Periodic Patterns in Two and
Three Dimensions. Ph.D.  Thesis, University of Houston (1990).

\bibitem{DG}   B.\ Dionne and M.\ Golubitsky. 
Planforms in two and three dimensions.  
{\em Z. Angew. Math. Phys.\/} {\bf 43} (1992) 36--62.

\bibitem{DSS}   B.\ Dionne, M.\ Silber  and A.C.\  Skeldon.  
Stability results for steady, spatially-periodic planforms.  
{\it Nonlinearity} {\bf 10} (1997) 321--353.


\bibitem{proctor} M.R.E. Proctor.
Planform selection by finite-amplitude thermal convection between poorly
conducting slabs.
{\it J.\ Fluid Mech.} (1981) 469--485.

\bibitem{SS} L.\ Shtilman and B.\ Sivashinsky.
Hexagonal structure of large scale Marangoni convection.
{\it Physica D} {\bf 52} (1991) 472--488.

\bibitem{ig} E.\ Ihrig and M.\ Golubitsky.
Pattern selection with O(3) symmetry.
{\it Physica D} {\bf 12} (1984) 1--33.

\end{thebibliography}
\end{document}